\documentclass[aps,11pt]{revtex4}
\usepackage[dvips]{graphicx}

\usepackage{amsmath}
\usepackage{bm}

\begin{document}

\preprint{}

\title{IS IT POSSIBLE TO PREDICT THE SIGN OF THE MATTER-ANTIMATTER ASYMMETRY IN THE UNIVERSE?}
\author{Pavel Fileviez P\'erez}
\email{fileviez@cftp.ist.utl.pt}
\affiliation{Centro de F{\'\i}sica Te\'orica de Part{\'\i}culas. 
Departamento de F{\'\i}sica. Instituto Superior T\'ecnico. 
Avenida Rovisco Pais,1\\ 1049-001 Lisboa, Portugal.}
\begin{abstract}
I investigate the possibility to define the sign of the leptonic 
asymmetry by the low energy parameters. It is shown that in the 
context of the minimal renormalizable $SO(10)$ model the sign of 
the matter-antimatter asymmetry can be defined by the 
leptonic mixing and masses in the case of Type II see-saw. 
\end{abstract}
\maketitle
The problem of the matter-antimatter asymmetry in the Universe is 
one of the most interesting problems in modern physics. There are 
many scenarios where we could explain the absence of 
antimatter in the Universe~\cite{Early}. Baryogenesis 
via leptogenesis is one of the most popular mechanisms where 
it is possible to predict the baryon asymmetry in a very 
simple way~\cite{Yanagida}.   

In this letter we investigate the possibility to define the sign 
of the baryon asymmetry in the Universe in the context of the 
$SO(10)$ models~\cite{SO(10)}. 

In minimal renormalizable $SO(10)$ the Higgs sector is composed by 
$10_H$ and $126_H$. In this model the Yukawa couplings for leptons 
are given by~\cite{Babu}:
\begin{eqnarray}
Y_N \ = \ U_{11}^H \ Y_{10} \ - \ 3 \ U_{21}^H \ Y_{126}\\
Y_E \ = \ D_{11}^H \ Y_{10} \ - \ 3 \ D_{21}^H \ Y_{126} \ = \ E_C^* \ Y_{E}^{diag} \ E^{\dagger}   
\end{eqnarray}
where $Y_N$ and $Y_E$ are the Dirac Yukawa coupling matrices for neutrinos 
and charged leptons, respectively; $U_{11}^H=v_{10}^u / v^{u}$, 
$D_{11}^H=v_{10}^d / v^{d}$, $U_{21}^H=v_{126}^u / v^{u}$, and 
$D_{21}^H=v_{126}^d / v^{d}$. The parameters $v_i$ are the 
expectation values entering in the theory. $E_C$ and $E$ are 
the matrices which diagonalize the Yukawa coupling matrix for 
charged leptons. Since the Yukawa couplings in this model 
are symmetric, $E_C = E K_3$. $E^{\dagger}N = K_e \ V_{PMNS}$, 
where $K_e$ and $K_3$ are matrices containing three CP violating phases, and 
$N$ is the matrix which diagonalize the Yukawa coupling matrix for left-handed neutrinos.

In those models the so-called Type II see-saw~\cite{see-saw} contribution 
for neutrino mass is given by:
\begin{equation}
M_{\nu}^{II} \ = \ Y_{126} \ v_L   
\end{equation}
while the mass of the right-handed neutrinos read as:
\begin{equation}
M_R \ = \ Y_{126} \ v_R
\end{equation}
with $v_L$ and $v_R$ the vacuum expectation values of the triplets 
$\Delta_L$ and $\Delta_R$, respectively (for more details see reference~\cite{Babu}). 
Now, using the above equations we can write $Y_N$ in the following way: 
\begin{equation}
Y_N \ = \  c_1 \ V_{PMNS}^T \ K \ Y_E^{diag} \ V_{PMNS} \ + \ c_2 \ M_{\nu}^{II}
\end{equation}
where $K= K_e^2 K_3^*$, while $c_1$ and $c_2$ are given by:
\begin{equation}
c_1=\frac{U_{11}^H}{D_{11}^H}
\end{equation}
\begin{equation}
c_2=3 \frac{(U_{11}^H D_{21}^H - U_{21}^H D_{11}^H)}{v_L D_{11}^H}
\end{equation}
Notice that we can choose the parameters $c_1$ and $c_2$ as real. Now, 
since the matrix $K_3$ is arbitrary or unphysical, we are allow to 
choose $K_3= K_e^2$, i.e. we can rotate the right-handed charged 
leptons in such way that we can satisfy this relation. 
You could think about the possibility that those phases contribute 
to any physical process. For example, in the most important 
prediction coming from grand unified theories, in the 
decay of the proton, the phases in $K_3$ do not appear in the 
different proton decay channels~\cite{PFP}. 
Therefore, working in the Type II see-saw limit the 
above expression for $Y_N$ read as:
\begin{equation}
\label{YN}
Y_N \ = \ c_1 \ V_{PMNS}^T \ Y_E^{diag} V_{PMNS} \ + \ c_2 \ M_{\nu}^{diag}
\end{equation}

Now let us investigate the implications for the different mechanisms of 
Baryogenesis via Leptogenesis~\cite{Yanagida} in the case of Type II see-saw.  

In reference~\cite{GoranHambye} the authors studied the different scenarios 
for leptogenesis in the context of left-right models. Let us analyze 
the cases when the Type II mechanism dominates. We have two 
cases~\cite{GoranHambye}: 
                                                     
\begin{itemize}

\item {\bf Case a)} $M_{N_K} << M_{\Delta_L}$ (take $k=1$ for the lightest right-handed neutrino)

In this case the lepton asymmetry is generated by the decays of the lightest 
right-handed neutrino. $M_{N_K}$ and $M_{\Delta_L}$ are the right-handed neutrino 
and triplet masses, respectively. The sign of the lepton-asymmetry in this case 
is defined by:   

\begin{equation}
\text{sign}(\epsilon^{\Delta_L}_{N_1})\ = \ - \text{sign}[ \ \text{Im} (Y_N \ M_{\nu}^{diag} \ Y_N^T)_{11} \ ]
\end{equation}

\item {\bf Case b)} $M_{\Delta_L} << M_{N_K}$

It is the so-called Triplet leptogenesis. Here the sign of the lepton asymmetry is given by:

\begin{equation}
\text{sign}(\epsilon_{\Delta_L}) \ = \ 
\text{sign} [ \ \text{Im} (Y_N^* \ M_{\nu}^{diag} \ Y_N^{\dagger})_{11} \ ] \ = \ \text{sign}(\epsilon^{\Delta_L}_{N_1})
\end{equation}
\end{itemize}

Now, since it has been shown before that in case of Type II see-saw, $Y_N$ is 
given by the Eq. (\ref{YN}). Therefore we can conclude that the sign of 
the lepton-asymmetry could be defined by matrix $V_{PMNS}$ and the leptonic 
masses. However, in those theories it is very difficult to predict the 
real parameters $c_1$ and $c_2$. The simplest way to predict the sign 
of the lepton asymmetry from the low energy parameters 
of the leptonic sector, is assuming that we have one massless neutrino, 
therefore the sign of the lepton asymmetry will be independent on the 
parameters $c_1$ and $c_2$. 
 
In future neutrino experiments we will know about all CP violating phases 
in the leptonic sector. Therefore we will able to understand much better 
quantitatively the connection between the sign of the baryon asymmetry and the 
low energy parameters in the context of the minimal renormalizable 
$SO(10)$ model.                                                  

The main conclusion of this Letter is that it is possible to find a direct 
connection between the sign of the matter-antimatter asymmetry and 
the low energy parameters of the leptonic sector in the context of the 
renormalizable minimal $SO(10)$ theory if the Type II see-saw term 
for neutrino masses dominates. The sign of the matter-antimatter asymmetry 
in the Universe can be defined by the low energy physical quantities. 

\begin{acknowledgments}
I would like to thank Gustavo C. Branco, R. Gonzalez Felipe, Thomas Hambye and 
Goran Senjanovi\'c for discussions and comments. 
I would like to thank G. Walsch for strong support.
\end{acknowledgments}


\end{document}